\documentstyle[12pt,epsf]{article}
\newcommand{\be}{\begin{equation}}
\newcommand{\ee}{\end{equation}}
\newcommand{\bea}{\begin{eqnarray}}
\newcommand{\eea}{\end{eqnarray}}
\newcommand{\ci}{\cite}
\newcommand{\bi}{\bibitem}
\newcommand{\nono}{\nonumber \\}

\newcommand{\dd}{\partial}

\def\dal{\,\lower0.3ex\vbox{\hrule\hbox{\vrulekern2pt\vbox{kern4ptkern4pt}
kern2pt\vrule}\hrule}\,}

\begin{document}

\vspace*{0.88truein}

\centerline{\bf Communication through an extra dimension}
\vspace{2 true cm} 
\centerline{G. K\"albermann}
\centerline{Soil and Water dept., 
Faculty of Agriculture}\
\baselineskip=10pt
\centerline{Hebrew University, Rehovot 76100, Israel}
\baselineskip=10pt
\vspace{2 true cm} 

\begin{abstract}
If our visible universe is considered a trapped shell
in a five-dimensional hyper-universe,
all matter in it may be connected by superluminal
signals traveling through the fifth dimension. 
Events in the shell are still causal, however, the 
propagation of signals proceeds at different velocities
depending on the fifth coordinate.

\end{abstract}


\newpage

\section{\sl The universe as finite size  shell}

Modern physics advocates for the existence of extra dimensions
besides the four we seem to live in. This assumption
creates enough room for a geometrical unification of
the different interactions between elementary particles
and their grouping in symmetry multiplets.

Our apparent blindness to these extra dimensions
is explained by resorting to the hypothesis
of compactification. It amounts to identifying all
the extra dimensions as essentially closed circles with
a minuscule (thereby undetectable) radius.\ci{review}
Such an explanation originated in the
work of Oskar Klein\ci{review}, and has arose
both support and antagonism over the years\ci{duff}.

On the other hand,
the possibility of the existence of large extra dimensions has been given
considerable attention recently. For compactified
large extra dimensions, it appears that, the experimentally safe
limit allows for them to be of
almost macroscopic size in submillimeter range.\ci{extra}
This concept was put forth for the first time by Antoniadis
in connection to the problem of supersymmetry breaking in string theory.
\ci{ant}

An even more intriguing possibility of our universe
as a shell has been called upon in order to
try and solve the hierarchy problem.\ci{hamed}
The latter approach seems to be consistent with Newton's law of gravitation
if the gravitons themselves are confined to the shell.\ci{randall}
The idea has also prompted cosmological investigations
and the scenario can be made consistent with the accepted models
of inflation and cosmological evolution.

Alternatively, the induced-matter approach of Wesson\ci{wesson} resorts to
extra dimensions that are not curled-up, in order
to generate matter fields.
Five-dimensional Einstein gravity of empty space is
used to generate our four-dimensional universe
both in its geometric (curvature) aspects and its
energy-momentum contents.
The very existence of mass and charge
of elementary particles is
claimed to be a consequence of the properties of a universe
with extra dimensions, a feature already noticed by Kaluza and Klein
in the early part of the century.\ci{review}
Mass could be related to the fifth
coordinate and charge to the momentum along it. 
The extra dimension shows up in the physical properties of
particles.

It seems that extra dimensions are slowly losing their
role as mere ancillary variables.

A few years ago,
several authors in the physics literature speculated about the possibility
that our universe is a thin shell in a larger dimensional hyper-universe.
In this approach, the extra dimensions are not curled-up,
hence their influence might be felt perhaps at an energy scale
lower than that needed to explore the
Planckian compactified dimensions.\ci{rubakov}

Visser\ci{visser} showed that the trapping of our universe in a thin
shell can be
implemented mathematically by using a large cosmological constant, originating
from a five-dimensional electromagnetic field energy.

Squires\ci{squires} later explicitly showed the trapping by using a cosmological
constant. Mater is represented by a scalar
field that becomes effectively confined to one dimension less than the
original space. The solution of Squires in three dimensions
is similar to the one obtained by Visser.\ci{visser}

More formal arguments for multidimensional spaces were given by 
Beciu\ci{beciu}.

Gogberashvili has shown that trapping also occurs when
a homogeneous background of electromagnetic energy
fills the shell\ci{gog}.
Gogberashvili's solution consists
of a metric conformal to flat space with a conformal factor
depending on the fifth dimension.
In this case, the stability of the shell is satisfied for a 
four dimensional space-time
immersed in a five dimensional manyfold only.

In the recent works aiming to solve
the hierarchy problem, our world is a three-brane embedded
in a larger dimensional universe.  The models resort to bulk a
cosmological constant and brane tensions (for more than one brane).
Consistency  with Newton's law even when there exist 
more than four noncompact dimensions is nevertheless maintained.\ci {randall}

In the present work we will show that, irrespectively of the mechanism of
trapping,
the very existence of the trapped shell,
implies a superluminal connection between all matter in the universe.\ci{hai}
In  other words, the speed of light increases the farther off-center
is the propagation of the signal, the center being defined
as the bottom of some potential well and identified
with our universe.
This effect may be regarded as time dilation, time runs slower
in the inner regions of the shell.
Causality is nevertheless maintained. Light cones will not 
tip enough in order to generate closed timelike curves, the whole shell
is causally connected.

As a consequence of the present results, we will find that
all matter and energy in our universe (the center of the shell), 
is tied up together in a
manner that usual normal signal propagation at the center of 
the shell would prohibit.
The speed of propagation of signals is not limited by the speed of light
measured on the four dimensional hypersurface.
If a signal penetrates inside the shell along the fifth dimension
it can leap forward huge distances in extremely short times as
compared to the time taken by signals propagating on
the hypersurface.
Similar findings have recently been described by Chung and Freese.\ci{freese}
They use more than one brane in a cosmological context,
the link between causally disconnected regions may then
be achived by means of signals proceeding through the extra dimension.
Hence, 
there is no need for inflation in order to solve the horizon problem.\ci{freese}
This is already evidenced in the results of ref.\ci{hai}.

In the next section we will examine the metric of five
dimensional space-time with an extra space-like dimension and
find the conditions for solutions to exist.
Section 3 will treat the
propagation of a signal along geodesics inside the shell and
provide some speculative thoughts implied by the present results.

\section{\sl Shell metric}

Before proceeding to analize the problem, we
show that the effect of hyperfast communication through an extra dimension
is already evident in the problem dealt by Visser.\ci{visser}

Visser used a cosmological constant modeled in terms
of an electric field in the fifth direction yielding a metric

\be\label{visser}
ds^2=cosh(E\xi)~dt^2-d\Omega_3^2-d\xi^2
\ee

where E is a constant , $d\Omega_3^2$ is the volume on the three brane
and $\xi$ is the extra dimension.

It is easy to integrate the geodesic equations the for metric above.
Consider a signal traveling in the $x,\xi$ direction
with initial speed at $\xi=0$, $\dot{\xi}_0=u,\dot{x}_0=v$.

We obtain

\bea\label{geovisser}
\dot{x}&=&v~cosh(E\xi)\nono
\dot{\xi}^2~&=&cosh(E\xi)-(1-u^2)~cosh^2(E\xi)
\eea

It is clear from the above equation that, the signal will
climb up the potential into the fifth dimension thereby accelerating 
beyond the normal speed of propagation at $\xi=0$ in an exponential manner.
The signal will stop at a point where $\dot{\xi}$ vanishes. Then will recede,
ocillate back and forth between endpoints propagating at great speed in
the $x$ direction inside the shell.
It will be quite hard to detect such a signal because it
spends almost all its time outside the center of the shell.

We will now examine the conditions that a
five-dimensional metric has to obey, in order to be appropriate for 
the trapped shell scenario, and show that even without a cosmological
constant, the effect persists.

For simplicity we take a isotropic and homogeneous four-dimensional 
hypersurface embedded along the fifth dimension. 

The line element consistent with this demand reads,

\be\label{element}
ds^2~= a(\xi,t)~dt^2-b(\xi,t)~d\xi^2-c(\xi,t)~d{\Omega_3}^2
\ee

where $a,b,c$ depend only on the
fifth coordinate $\xi$ and the cosmic time $t$.

The above ansatz corresponds
to a hyperspherically symmetric solution of five dimensional Einstein
gravity, provided Einstein's gravity still works
in the hyper-universe, as one might hope. We still consider
space-time inside the shell as a Riemannian manifold.

Define the center of the hypersurface by $\xi=0$, such that
$a(0,t)=1,~b(0,t)=A(t)$, with $A$ the scale parameter of our expanding
universe. This coordinate choice 
defines what is understood by the cosmic time on the shell, at $\xi=0$.

The shell is allowed to have a finite -albeit minuscule- width.
We have chosen a flat four-dimensional metric. Distances in the shell
at fixed $\xi$ are not limited. We could have chosen a curved shell, however,
locally, the effect of cosmic curvature should be negligible.
Moreover, a similar calculation with a closed (or open) universe
yields analogous results.
We have also chosen the signature of the extra dimension to be space-like.
Time-like extra dimensions lead to tachyonic gravitons and other exotic
phenomena.\ci{dvali}

We now simplify the problem by taking
all the metric functions to be time independent, namely we choose
$t=t_0$, the present epoch.
We do so because the scale function of conventional Friedmann-Robertson-Walker
expansion is the inverse Hubble constant, that is
extremely large
as compared to the supposed thickness of the shell.
We are not interested in cosmological evolution, but, instead in the effects
of communication through the extra dimension in our present time.
The cosmological aspect may be found in ref.\ci{freese}
Time evolution of the shell 
is not important for the time scales that will be involved in the present
investigation.

Inserting the ansatz of eq.(\ref{element}) into the Einstein tensor we
find

\bea\label{tensor}
G_{00}&=&\frac{3a}{4~c^2~b}~(2~b''~c-b'~c')\nono
G_{ii}&=&-1/4~(2~b~a'b'ca-a^2~b'^2~c+4~a^2~b''bc-\nono
&-&2~a^2~b'~c'~b+2~b^2~a''a~c-b^2~a'^2~c-b^2~a'~c'~a\bigg)/(a^2~b~c^2)\nono
G_{44}&=&-\frac{3~b'}{4~a~b^2}~(a~b)'
\eea

primes denoting derivatives with respect to $\xi$ and $ii=11,22,33$.

Regardless of the trapping mechanism (cosmological constant, negative
energy, background fields)
the function $a$ related to the gravitational potential,
may be expanded around the equilibrium (or
quasiequilibrium, due to the time dependence) radius $\xi=0$.
At fixed time, the expansion has to consist of a constant and a quadratic term. 
The linear term has to vanish due to the
equilibrium condition, otherwise, all the mass in the universe would suffer
a constant force that will eventually drag it to a new equilibrium position
and we can always call this new position $\xi=0$. 

Assuming, that the extension of the shell into the fifth dimension is
tiny as compared to any reasonable macroscopic scale length -as it should be
if its effect remains undetected in the dynamics of bodies-,
our approximation will be valid. 

Hence we have

\be\label{func}
a(\xi,t)=~a_0+k~\frac{\xi^2}{2} 
\ee

Space-time is flat at the
center of the shell $\xi=0$. In order to connect to the Minkowski
metric of our flat world, we take $a_0=1$, without loss of
generality.
The second assumption is that the solution we
are looking for is surrounded by some agent.
Hence it is an external approximately vacuum solution. 
External in the sense, that the brane is an almost empty region of space
sandwiched between hypervolumes filled with some unknown
substance.

As in the Schwartzschild case we will then have $ c(\xi)=a(\xi)^{-1}$.
Such an ansatz simplifies the equations radically, and
can be viewed as a redefinition of the coordinate $\xi$.

A solution in and around the shell location
may be obtained in several manners. 
(Recall we are not demanding an
exact solution to all orders in $\xi$.)

One such solution obtains with $b(\xi)=e^{-\beta\xi^2\over2}$.
For which , $G_{00}\approx-\frac{3}{2}~\beta
+O(\xi^2)$, the other components of the Einstein tensor
vanishing provided that $\beta=k$.

The energy-momentum tensor implied by eq.(\ref{tensor}) has then 
a vanishing pressure and an energy density that is nonzero, a cloud
of dust.

One should bear in mind that the dust we need in order to
confine the shell, is not the matter dust of our world.
The latter is a hundred of orders of magnitude less dense.
This is also the case when one uses a cosmological constant.
Particle physics teaches us that the cosmological
constant should be of the order of a density 
of $\Lambda\approx TeV/fm^3$, while, if at all needed for the purposes
of cosmology, we should have a cosmological constant 120 orders
of magnitude smaller.
The fact that both the cosmological constant and the energy density 
needed in order to confine the shell are enormous, 
in cosmological terms, is quite disturbing.
However, these parameters 
are merely hiding our ignorance concerning the kind
of matter-energy that exists beyond our visible world.
Moreover, the huge cosmological constant
used in the literature to generate the brane, or, in the
present case, the huge energy density, serve only
for that specific purpose. We are using the
vacuum energy of particle physics to stabilize the brane.
In some sense this is an economical approach.

The difference between our solution and those that use
a cosmological constant\ci{visser}, is that the present
solution is analogous to the so called quintessece 
models\ci{di pietro},presently adduced in order to 
explain the acceleration (negative deceleration) parameter
of the universe. In those models the pressure is smaller
(and perhaps even zero) as compared to the energy density.
In any event, we will find the same hyperfast travel effect
as with the cosmological constant, supporting the claim that
it is quite a general feature of trapped shells.

The basic parameter that determines the curvature of the shell
is $k$. It should be large enough, for the
thickness of the shell to be smaller than the radii of nuclei,
smaller than the deep inelastic scattering scale of the experiments
carried until now, otherwise, 
its imprint should have become visible by now.
Recent works on submillimeter extra dimensions,
seem to indicate that for the compactified scenario, the
upper limits on the size of these dimensions could be much higher,
even macroscopic.\ci{extra} We here opt, however, for a much more
conservative approach and assume the shell size parameter is
much smaller. The results are in any event 
independent of the choice, but are much more evident
the thinner the shell.

Summing up, we have a metric with

\bea\label{sum}
g_{00}&=&1~+~k\xi^2\nono
g_{ii}&=&-e^{-k\xi^2/2}\nono
g_{44}&=&-\frac{1}{1+k~\xi^2}\nono
\eea

Let us now consider signal propagation.
We will here use an extremely schematic approach in order
to estimate the effect of the shell thickness
on the signals. A geodesic
method will be exploited in the next section. 
For this purpose we assume
propagation along a path that goes in the $\xi$ direction, then
across it and back.

Our measure of time in the shell is the cosmic time $t$,
and we refer every process to it.
Suppose a massless field signal whose speed in the shell 
is the velocity of light $c=1$ in our units,
starts traveling from $\xi=0$, the center of the shell to some fixed $\xi$
inside it, then travels a distance $L$ 
at fixed $\xi$ and returns from $\xi$ to $\xi=0$ back to a
point at a coordinate distance $L$ from the initial point. 
Signals emitted in a direction 
at an angle to the hypersurface, scatter
inside it, or may be reflected at some hypothetical edge. 
The question of the dynamics of radiation inside the shell
is here relevant, especially if photons, gravitons, etc., become
massive. 
We here confine ourselves to a kinematical approach. A rough estimate
of the superluminal effect is obtained below, while
a geodesic approach is taken up in the next section.

Using three dimensional spherical
coordinates, $L$ along the radial distance in three-dimensional
space, and for fixed angles, we find

\be\label{elem}
ds^2=(1+k~\xi^2)dt^2-\frac{1}{(1+k~\xi^2)}~d\xi^2-e^{-k~\xi^2/2}~dl^2
\ee

Our choice of $k$ will be of the order of $R_{GUT}^{-2}$, where
$R_{GUT}\approx10^{-31}m$ stands for the grand unified theories scale. 
We do this because we want to encompass democratically all the known
interactions, and, in order to avoid quantum gravity
effects that will enter at much smaller scales of the order
of $R_{Planck}\approx~10^{-35}m$.

However, any microscopic scale will be a viable choice. We take this
value for the sake of exemplification

Hence

\begin{equation}\label{t}
t=~2\int_0^\xi{\frac{d\xi}{1+k~\xi^2}}~+~\int_0^L{\frac{dl~e^{-k~\xi^2/4}}
{\sqrt{1+k~\xi^2}}}
\approx~\frac{L~e^{-k~\xi^2/4}}{\sqrt{1+k~\xi^2}}
\end{equation}

Where we neglected the first integral
because it is of the order of $t\approx10^{-39}sec$.

If the signal climbs up the harmonic potential and back far enough
in $\xi$, the coordinate time becomes negligible.

With $\xi= 15~R_{GUT}\approx 1.5~10^{-30}m$, several times
the radius of curvature of the shell, and $L= 100 Mpc$, 
the time taken by radiation to traverse this cosmic distance is
$t=2.5~10^{-10}sec$. A ridiculously small time 
as compared to the 326 Million years 
needed for the light to traverse this distance along the direction 
$\xi=0$.

Due to the crudeness of the approximations used in order
to derive the above result, we should not attach too much rigor to
the actual numbers. The effect is, nevertheless, evident.

The same results are obtained in the eikonal approximation for
a scalar field propagating on the shell. This was indeed
checked by using a conformal factor in the metric and identifying
the departure of the factor from the value of one as the 
propagating scalar field.

\section{\sl Geodesic propagation of signals in the shell}

In the previous section we considered a simplified
scenario of signal propagating along straight lines inside the
shell in order to estimate the effect of superluminal communication
between different points in the embedded universe.
However, straight lines are not the correct paths of
either massive or massless particles.
We here proceed one step forward and study the propagation
of a signal in the shell using the metric found in the previous section
as a background.

Consider the propagation of a signal in the $\xi,x$ plane, where
$x$ is a coordinate along the shell at $\xi=0$.
The geodesic equations for the $t,\xi,x$ coordinates become

\bea\label{geodesics}
t''+\frac{2~k~\xi}{1+k~\xi^2}~t'~\xi'=0\nono
x''-\beta\xi~x'~\xi'=0\nono
\xi''+t'^2~k~\xi~(1+k~\xi^2)-\frac{k\xi}{1+k\xi^2}\xi'^2+
\frac{\beta\xi}{2}(1+k~\xi^2)~e^{-\frac{\beta~\xi^2}{2}}x'^2=0
\eea

Where the primes represent here derivatives with respect to the
parameter of the geodesic equations.

The first equation of the set above may be solved by $t'=(1+k~\xi^2)^{-1}$,
fixing the parameter of the geodesics to be the time $t$ at $\xi=0$
up to an irrelevant constant.
The motion along $x$ is then determined by $\frac{\dd x}{\dd t}=
C~(1+k~\xi^2)~e^{\frac{\beta\xi^2}{2}}$, 
with $C$ a constant of integration determined by the projection
of the initial velocity on the $x$ axis.
The equation for the $\xi$ coordinate then becomes (the equation may
be also obtained from variation of the path distance $-\int~ds$)

\be\label{xieq}
\ddot{\xi}-\frac{3k\xi\dot{\xi}^2}{1+k\xi^2}+k\xi(1+k\xi^2)+C^2\beta
\xi/2~(1+k\xi)^3~e^{\beta\xi^2/2}=0
\ee

The equation has no analytical solution. If one neglects the nonlinear
terms, the solution becomes a harmonic function as expected.
However, these terms are extremely important for $\xi$ large enough.
The equation for $\xi$ has to be solved numerically. Once $\xi$ is found
the displacement of the particle along the shell may be found.

We use $k=\beta=1$ in
units of $\frac{1}{R_{GUT}^2}$, and a signal
impinging at an angle $\alpha$ on the shell with velocity components
$\dot{x}_0,\dot{\xi}_0$.
It is found that regardless
of the angle of incidence, the particle oscillates back and
forth across the shell with a frequency of the
order of $\sqrt{k}$.
The long-time behavior of the signal requires an
extremely fine grid in order to prevent roundoff errors.
Long time is still a minute time as it is measured
GUT time units.
For example, a signal with initial speed $0.9c$ at an angle of 45 degrees
with respect to the $\xi=0$ axis has traversed a distance $L= 4~c~t$
after t=7000 time units $\approx 2~10^{-36}$ sec,
It is clear that we recover the results of the simplistic calculation
of the previous section.
The signal travels superluminally cosmic distances. 
The gain increases as time goes on.

The possibility of disappearance of signals into the brane and
their reappearance in some other part of the universe is
quite perplexing.
One immediate consequence is that, the flux coming from a source
decreases faster than $\frac{1}{r^2}$, flux is lost.
The caveat is that we used a geodesic approach. It  is
not all certain that this is a sound procedure,
especially because particles cease to appear
pointlike as compared to the size of the brane in the
fifth dimension.
In order to give some more credibility to the
present results, we need a dynamical theory of fields
and particles. However, our clues coming from
our experience on the brane seem a bit irrelevant in this respect.
The use of a geodesic equation despite its limitations is, in some
sense, the least biased approach. Any dynamical model, based
upon field theory as we know it may fail
to describe the physics of the entities in a larger number
of dimensions. We need some experimental input
in order to proceed. In the absence of such evidence
it is preferable to leave the situation as is.

For observers far from the source, the luminosity
of a source and therefore its mass, are underestimated.
If taken seriously, the present effect has to be included
in the calibration of radiating sources both near and far.

Perhaps even for a nearby radiating source we
are witnessing only the 'tip of the iceberg' of its energy output.
It is quite evident that gravitational radiation
will also be scattered all over the shell. Wandering radiation
and particles inside the shell create a background noise
present in the whole universe, perhaps serving as the very source
of the energy density used above in order to maintain the
shell's stability.

If a photon does indeed penetrate the thicket of the shell
then it will become red shifted and blue shifted enormously, but will arrive
at its destination with the same frequency as emitted.
However, the very existence of
photons inside the shell has to be questioned.

In conventional Kaluza-Klein
approaches, the electromagnetic field (potential) is considered
a four-vector from the onset, there is no electromagnetic
potential in the extra direction.
However, if the
extra dimension is a physical one and not curled-up, there is
a-priori no clear differentiation between, say, the middle
of the shell and, nearby positions along the fifth direction.
The presence of the electromagnetic field in the
fifth direction has to be considered also, at least classically.
Moreover, if one relaxes the condition of
cylindricity, there arise fields that propagate along
the extra dimension, and correspond
to the electromagnetic fields in the shell.

The cylinder condition arises in compactified models
when one uses the lowest order mode of the metric
expanded in terms of Fourier amplitudes.
The mechanism of compactification is still unclear.
The compactification ansatz was devised by Klein
to justify the cylinder condition of Kaluza. Klein himself
restricted $g_{44}$ to be a constant. Later \ci{review}, it
was found that such an assumption is inconsistent. It gives
an extra equation for the electromagnetic field that is not
satisfied by real fields. The lesson we learn form Klein's
restriction and its failure is that one should not constrain
the fields beforehand.

If indeed light might be lost in the depth of the shell,
we have an alternative explanation to the missing matter
(or missing light) problem. In this case, the velocities of galactic dust 
clouds 
are determined by the gravitational masses present in the galaxies, but, the
light emitted by the galaxies underestimates the mass, flux is lost.
The cosmic distance ladder has to be recalibrated in order
to account -at least statistically- for the missing light.
Moreover, it could be that our universe-shell is 
cracked outside the regions where galaxies reside.
The velocity of orbiting clouds is then determined
not only by the positive mass inside the galaxies,
but, by the invading mass (presumably negative)
that comes from the regions external to the brane.

Radiation emitted into the shell
is detected almost instantaneously by particles far away in the universe in
a random-like manner.
The fact that we do not violate causality and locality is because
both are distorted enormously by the potential of the shell.
This in turn might have some bearing to the nonlocality witnessed
in quantum mechanics. Instead of having alternative hidden-variable
theories we could think about hidden-dimension theories.
Another aspect of the superluminal propagation is that there
is no need for an inflationary era of cosmological evolution.\ci{freese}
The homogeneity and isotropy of the universe, evidenced
by the microwave background, can be achieved by means
of the same mechanism. Far away points in the universe
do not have to wait until they enter each other's 
horizon, they are communicated at all times through
the extra dimension.
Differing from ref.\ci{freese}, the brane we use is of
finite size. If it were of zero width
it would be hard to view a signal (not only photon),
as a messenger. (see remark below eq.(\ref{geovisser})).
However, if matter is extended along the
fifth dimension too, as it should if if the brane is of finite size,
then there is no such problem, matter will detect
the signal even if it does not transite through the middle of the shell.

The model addresses what appears to be action at a distance
in terms of ultrafast communication. 
The incessant bombardment by
ultrafast radiation that fills the universe's shell,
may be also related to the zeropoint field of quantum theory
because of its random character with no definite temperature
signature.

\end{document}